# All-optically induced currents resulting from frequency modulated coherent polarization

**Shekhar Priyadarshi, Klaus Pierz, and Mark Bieler**

*Physikalisch-Technische Bundesanstalt, 38116 Braunschweig, Germany*

We employ polarization-shaped ultrafast optical pulses to generate photocurrents which only arise if the optically induced coherent polarization is frequency modulated. This frequency modulation is obtained via detuned excitation of light-hole excitons in (110)-oriented GaAs quantum wells. The observed photocurrents vanish for resonant excitation of excitons and reverse their direction with a change of the sign of detuning. Moreover, the currents do not exist for continuous-wave excitation. Our work reveals the existence of a new class of photocurrents and visualizes the complexity of current response tensors. This is helpful for the better understanding of optically induced microscopic transport in semiconductors.

The optically induced coherent polarization plays a crucial role in linear and nonlinear light-matter interaction. In semiconductors a slowly dephasing coherent polarization results in many interesting phenomena such as quantum beats, photon echoes, and Bloch oscillations, which were previously observed in four-wave mixing and photocurrent experiments.[1-9] All these effects show a nonlinear dependence on the optical excitation fields. Focusing on nonlinear second-order effects and above-bandgap excitation of semiconductors, difference frequency mixing between the coherent polarization and the optical excitation field may lead to injection (IC), symmetric shift (SSC), and antisymmetric shift (ASC) currents.[10-14] On a macroscopic level these currents can be linked to 3$^{rd}$-rank tensors and have a certain dependence on the optical phase ($\varphi$) between the two optical excitation fields.[15]

The IC, which results from a spin-selective excitation of semiconductors leading to asymmetrically distributed carriers in momentum space, is associated with response tensors which are imaginary and antisymmetric in their last two Cartesian indices.[13,16] The SSC and ASC result from a real-space shift of carriers during interband excitation. The tensors associated with the SSC and ASC are both real, yet the SSC tensors are symmetric while the ASC tensors are antisymmetric in their last two Cartesian indices.[13,14,17] Due to the required optical phase $\varphi$ between the excitation fields the IC is also known as circular-photogalvanic current with a $\sin(\varphi)$ dependence while the shift current is also referred to as linear-photogalvanic current with a $\cos(\varphi)$ dependence.[13]

In this letter we report on the observation of new shift- and injection-current components which only arise if the optically induced coherent polarization is frequency modulated. Such frequency modulation is obtained from detuned excitation of light-hole excitons (lhX) in a GaAs quantum well (QW) and thus extends our previous work [14], which focused on resonant excitation only. The new current components are accessed using various ultrafast polarization-shaped optical pulses. Interestingly the direction of these currents is reversed with a change in the sign of detuning of the optical photon energy with respect to the exciton resonance. All experimental results are well explained using a simple model based on the Bloch equations in the length gauge.[13,17] We note that previous studies in which the frequency modulated coherent polarization has been shown to play an important role addressed different physical effects.[6]

The experiments are performed on an (110)-oriented GaAs/Al$_{0.3}$Ga$_{0.7}$As quantum well (QW) sample comprising 40 periods of 5 nm wide wells. We employ the (110) orientation mainly due to experimental reasons (current flow in the plane of the QW for normally incident radiation) and like to emphasize that our results are valid for any non-centrosymmetric crystal. This is because the frequency modulated coherent polarization can be induced in any crystal class via non-resonant excitation of excitons. The optical excitation is obtained from a pair of time-delayed, unchirped, and orthogonally-polarized femtosecond laser pulses (140 fs temporal and 5.2 nm spectral width). Their center wavelength can be tuned across the photon energy of the lhX transition of the sample. The peak intensity of each individual beam is kept at 3.5 MW/cm$^2$ resulting in a carrier density of ~1×10$^{10}$/cm$^2$ per pulse. The electric fields of the two excitation beams are aligned along the x = [001] and y = [1$\bar{1}$0] in-plane crystallographic directions of the QW sample. In this geometry we excite a superposition of IC, SSC, and ASC along the y direction employing the yxy and yyx tensor elements with the strengths of the currents being dependent on φ. The currents along y are detected by measuring the simultaneously emitted terahertz (THz) radiation via electro-optic sampling with a time-delayed probe beam.[18] All experiments are done at room temperature.

In the following we present THz interferograms (TI), showing the THz amplitude at a certain probe-beam delay versus time delay τ = φ/ω$_c$ between the two excitation pulses. These TIs oscillate with the optical carrier frequency (ω$_c$), but have certain slowly-varying features that are, as shown below, of interest for us. In order to suppress the IC in these measurements we have selected the delay of the probe beam at τ = 0 such that the IC amplitude is negligible as compared to the SSC amplitude, see inset of Fig. 1. To this end, we employ the fact that the THz traces of shift and injection currents obtained from excitation of the lhX differ from each other.[11] During the measurements of the TIs we have simultaneously adjusted the probe-beam delay by τ/2 to account for the change of τ. As seen below the adopted variation of the probe beam is important, since it provides even (odd) TIs with respect to τ for symmetric (antisymmetric) tensor elements.[14] In order to calibrate φ between the two excitation pulses we have simultaneously measured their electric-field correlation (EFC) using an ellipsometric technique.[19] For the discussion of the experimental results we employ the phase differences between the TI and the EFC. These phase differences are equal to even and odd multiples of π/2 for a scenario in which the current being at the origin of the measured TI has a cos(φ) and sin(φ) dependence, respectively.

In Figs. 1(a), (b), and (c) are plotted TIs obtained from excitation of the n=1 lhX with slightly different photon energies corresponding to resonant and non-resonant excitation conditions: (a) ~4 meV below resonance, (b) at resonance, and (c) ~4 meV above resonance. In Fig.1 (b) perfect waists occur on either side of $\tau = 0$. These waists appear due to counteracting SSCs that have different amplitudes and dephasing rates. While the SSC obtained from excitation of lhXs has a small amplitude at $\tau = 0$ and experiences rather slow dephasing, the SSC obtained from excitation of the heavy-hole dominated continuum carriers[20] has a larger amplitude at $\tau = 0$ but experiences faster dephasing. At a certain time delay $\tau$ the strengths of these two currents are equal which results in the observed sharp waists.[10, 14] Moreover, the asymmetry of the TI has been explained by constructive and destructive interference between the SSC and ASC, both resulting from excitation of the lhX, on the right- and left-hand side, respectively, of the TI.[14] From Figs. 1 (a) and (c) it becomes evident that the waists already disappear for very small detuning, yet the asymmetry of the TI is retained. The phase differences of the corresponding TIs are plotted in Figs. 1 (d), (e), and (f). For resonant excitation of the lhX, the phase difference shows an abrupt jump of $\pm\pi$ at the time delay corresponding to the waists of the TI, see Fig. 1 (e). The phase jumps have the same origin as the waists in the TI. However, for non-resonant excitation the phase differences show a gradual, close-to-linear variation with the sign of the slope being dependent on the sign of detuning, see Figs. 1 (d) and (f). The disappearance of the waists reveals the presence of currents with both $\cos(\varphi)$ and $\sin(\varphi)$ dependences since these two orthogonal functions cannot cancel each other. Moreover, the gradually varying phase difference shows that the ratio of currents with $\cos(\varphi)$ and $\sin(\varphi)$ dependences varies as a function of $\tau$. Thus for non-resonant excitation of discrete transitions additional current components evolve that have a $\sin(\varphi)$ dependence as compared to the $\cos(\varphi)$ dependence of the SSC and ASC.

Since Fig. 1 shows a clear dependence of the slope of the phase difference on excitation photon energy, we have performed additional measurements in an extended photon-energy range. The result is shown in Fig. 2 (a), where we plot the slope of the phase difference near $\tau = 0$ versus photon energy. From this plot it becomes evident that the slope of the phase-difference switches its sign at the lhX, where it has absolute values of ~1 rad/100 fs. The absolute value of the phase-difference slope approaches zero if the detuning from the lhX is increased. For larger detuning the population of non-resonantly excited lhX decreases and becomes negligible with respect to continuum carriers. Hence, at such photon energies the current response is similar to the current response obtained from continuum excitation with a phase-difference slope equal to zero. This result adds confidence in our assumption that the disappearance of the waists and the gradually varying phase difference results from non-resonantly excited lhX transitions. We note that similar results are obtained for measurements in which heavy-hole exciton transitions are excited. Yet we restrict our discussion to the lhX, since at the hhX the IC cannot be suppressed effectively due to similar shapes of the SSC and IC which in turn complicates the data analysis.

To analyze our experimental results we adopt a simple two-level model based on the Bloch equations in the length gauge. [13, 14, 17] Yet, in contrast to our previous work[14] the model accounts for non-resonant excitation, too. As described in Ref.[13] the injection and shift currents that are induced by optical interband excitation can be written as: $J_{IC}^y(t) = e \int \frac{d\mathbf{k}}{8\pi^3} \sum_n v_{nn,\mathbf{k}}^y \rho_{nn,\mathbf{k}}^{(2)}$ and $J_{SC}^y(t) = -\frac{e^2}{\hbar} \int \frac{d\mathbf{k}}{8\pi^3} \sum_{n,m} \rho_{nm,\mathbf{k}}^{(1)} r_{mn;y} \cdot \mathbf{E}$, respectively. Here, $|n\rangle$ and $|m\rangle$ are the two bands involved in the excitation process, $v_{nn}^y$ is the group velocity in the $n^{th}$ band, $r_{mn}$ is the interband transition dipole moment (TDM), and $r_{mn;y}$ is the generalized derivative of the TDM, see references[13, 14, 17]. The optical excitation field is denoted by $\mathbf{E}(t) = \mathbf{E}_{env}(t) \exp(-i\omega_c t) + c.c.$ with $\mathbf{E}_{env}(t)$ and $\omega_c$ being the electric-field envelope and the carrier frequency, respectively. We obtain expressions for the density matrix elements $\rho_{nm,\mathbf{k}}^{(1)}$ and $\rho_{nn,\mathbf{k}}^{(2)}$ from first- and second-order perturbation solutions, respectively, of the Bloch equations and assume a time-delayed pulse pair as excitation field $\mathbf{E}_{env}(t) = \hat{x} E_{env}^x(t-\tau) \exp(i\omega_c \tau) + \hat{y} E_{env}^y(t)$. Inserting $\rho_{nm,\mathbf{k}}^{(1)}$ and $\rho_{nn,\mathbf{k}}^{(2)}$ into the injection- and shift-current equations we can write after some simple calculations:

$$J_{IC}^y(t) + \frac{J_{IC}^y}{T_s} = \int \frac{d\mathbf{k}}{8\pi^3} i\eta^{yxy} \big[\{f_{env,R}^x(t-\tau)E_{env}^y(t) + f_{env,R}^y(t)E_{env}^x(t-\tau)\}\sin(\omega_c\tau) + \{f_{env,I}^x(t-\tau)E_{env}^y(t) - f_{env,I}^y(t)E_{env}^x(t-\tau)\}\cos(\omega_c\tau)\big] \quad (1)$$

$$J_{SSC}^y(t) = \int \frac{d\mathbf{k}}{8\pi^3} \sigma_s^{yxy} \big[\{f_{env,R}^x(t-\tau)E_{env}^y(t) + f_{env,R}^y(t)E_{env}^x(t-\tau)\}\cos(\omega_c\tau) - \{f_{env,I}^x(t-\tau)E_{env}^y(t) - f_{env,I}^y(t)E_{env}^x(t-\tau)\}\sin(\omega_c\tau)\big] \quad (2)$$

$$J_{ASC}^y(t) = \int \frac{d\mathbf{k}}{8\pi^3} \sigma_a^{yxy} \big[\{f_{env,R}^x(t-\tau)E_{env}^y(t) - f_{env,R}^y(t)E_{env}^x(t-\tau)\}\cos(\omega_c\tau) - \{f_{env,I}^x(t-\tau)E_{env}^y(t) + f_{env,I}^y(t)E_{env}^x(t-\tau)\}\sin(\omega_c\tau)\big] \quad (3)$$

Here $T_s$ is the intraband momentum scattering time,[18] and $f_{env,R}^{x,y}(t)$ and $f_{env,I}^{x,y}(t)$ are the real and imaginary parts, respectively, of the modified electric field envelope $f_{env}^{x,y}(t) = \int_{-\infty}^{t} E_{env}^{x,y}(t') \exp\left(-(\gamma + i\delta)(t-t')\right) dt' = E_{env}^{x,y}(t)(\gamma + i\delta)^{-1} - \dot{E}_{env}^{x,y}(t)(\gamma + i\delta)^{-2} +$ $\ddot{E}_{env}^{x,y}(t)(\gamma + i\delta)^{-3} - \cdots$. The dephasing rate is denoted by $\gamma$ and $\delta$ is the detuning. The IC tensor element $\eta^{yxy} = i\frac{4e^3}{\hbar^2}\Delta_{12,\mathbf{k}}^y Im(r_{12}^x r_{21}^y) = -\eta^{yyx}$ is imaginary and antisymmetric in the last two Cartesian indices, with $\Delta_{12,\mathbf{k}}^y = v_{11,\mathbf{k}}^y - v_{22,\mathbf{k}}^y$. The real quantities $\sigma_s^{yxy} = \frac{2e^3}{\hbar^2}Im(r_{12}^x r_{21;y}^y - r_{21}^y r_{12;y}^x) = \sigma_s^{yyx}$ and $\sigma_a^{yxy} = \frac{2e^3}{\hbar^2}Im\left(\frac{\partial r_{12}^x r_{21}^y}{\partial k^y}\right) = -\sigma_a^{yyx}$ denote the SSC and ASC tensor elements respectively, with $\sigma_s^{yxy}$ being symmetric and $\sigma_a^{yxy}$ being antisymmetric in the last two Cartesian indices.

Before presenting results obtained with this model some remarks are necessary. First, in equations (1), (2), and (3) both tensor elements yxy and yyx are included. Second, the second term in each of the above equations which includes $f_{env,I}^{x,y}(t)$ vanishes for resonant excitation and is negligibly small for excitation of carriers with very fast dephasing ($\gamma \gg \delta$). The latter will be the case for excitation of continuum transitions at room temperature. However, $f_{env,I}^{x,y}(t)$ exists if the coherent polarization is frequency modulated. Here we like to note that in addition to using non-resonant excitation of discrete transitions, frequency modulation of the coherent polarization may also be obtained through chirped-pulse excitation. Third, in bulk GaAs samples with a $T_d$ point group symmetry only the first term of $J_{SSC}^y(t)$ survives at room-temperature due to fast dephasing of carriers and additional symmetry considerations.[13] Fourth, no local-field correction has been taken into account since up to the second-order perturbation solution the local-field correction only shifts the resonance frequency slightly. Its exact value is not critical for our phenomenological model.[21]

From the model outlined above we see that each current (IC, SSC, ASC) can be divided into a resonant component (R, first term of Eqs. (1), (2), (3)), which is maximized for resonant excitation, and a non-resonant component (NR, second term of Eqs. (1), (2), (3)), which vanishes for resonant excitation (where the frequency of the coherent polarization is not modulated). These current components are expressed through corresponding subscripts R and NR. The SSC$_R$ is induced by a linearly polarized optical pulse, persists for continuous-wave (CW) excitation, and has a cos($\varphi$) dependence on the optical phase. The SSC$_{NR}$ is accessed using an elliptically polarized optical pulse whose principle axes are parallel to the x and y directions and their length ratio is initially greater than 1, reduces to 1 in the middle of the pulse, and successively becomes less than 1 at the end of the pulse. This current vanishes for CW excitation, and has a sin($\varphi$) dependence. The ASC$_R$ and IC$_{NR}$ are accessed using a linearly polarized optical pulse whose linear polarization slowly rotates in time, vanish for CW excitation, and have a cos(φ) dependence. Finally, the IC$_R$ and ASC$_{NR}$ are induced by a circularly polarized optical pulse, exist for CW excitation, and have a sin(φ) dependence.

The expected TIs of the different current components are schematically plotted in Fig. 2(b)-(e). Apart from the sin(φ) and cos(φ) dependences the various current components may lead to an 'even envelope' of the TI which is maximum at τ = 0 and decays with a rate related to dephasing or an 'odd envelope' of the TI which is zero at τ = 0 and has a maximum and minimum at some finite τ. The name 'even (odd) envelope' is used because the envelope is an even (odd) function of τ. It is the product of the sin(φ) or cos(φ) functions and the envelope which determines the symmetry of the TI. In general all current components with symmetric (antisymmetric) tensor elements lead to even (odd) TIs. Thus, for non-resonant excitation of discrete transitions a mixture of different TI envelopes with different dependences on φ occurs for the IC, SSC, and ASC. It is this mixture that is required to explain the observed gradual variation of the phase difference and the disappearance of the waists.

Simulation results based upon our simple model are shown in Fig. 3. We denote the SSC response tensor resulting from heavy-hole dominated continuum excitation in the bulk substrate and the QW region by $\sigma_b^{yxy}$. The response tensors of the SSC, ASC, and IC resulting from the lhX excitation in the QW region are denoted by, $\sigma_s^{yxy}$, $\sigma_a^{yxy}$ and $\eta^{yxy}$, respectively. We have assumed a relative strength of the tensor elements $\sigma_b^{yxy}:\sigma_s^{yxy}:\sigma_a^{yxy}:\eta^{yxy}$ being equal to 1:-0.33:0.046:0.046$i$. These values already account for the optically induced population difference between the QW region and the bulk substrate. The dephasing rates of the lhX and the continuum carriers have been assumed to be equal to 200 fs and 80 fs, respectively.[10, 14, 18] With this assumption we calculate THz traces for different current components at different τ and also considered a transfer function accounting for the EO detection.[19, 22] For τ = 0 we choose the temporal position $t_p$ where the THz trace obtained from the IC$_R$ is negligible as compared to the THz trace obtained from the SSC$_R$. To construct the TIs we then plot the overall THz amplitude at $t_p$ + τ/2 versus τ. Figure 3 nicely reproduces all of the experimental observations, in particular, the asymmetry of the TIs and the gradually varying phase difference.

Despite the excellent agreement between experiment and simulation, a more detailed discussion about the possible distinction between the non-resonant current components and their contribution to the measured and calculated TIs is required. As mentioned earlier, current components with a sin(φ) dependence, i.e., $IC_R$, $ASC_{NR}$, or $SSC_{NR}$ are required to explain the disappearance of the TI waists and the gradually varying phase differences. The $IC_R$ has been eliminated from the interferometric measurements by a proper choice of the probe delay, see discussion above. Moreover, our simulations show that in a scenario in which the SSC is considerably stronger than the ASC, inclusion of the $ASC_{NR}$ will lead to a phase-difference offset at $\tau = 0$ which is not appreciable in our measurements, see Fig. 1 (d) and (f). Hence, we conclude that the disappearance of TI waists and the gradually varying phase difference mainly result from the $SSC_{NR}$. This current component does not lead to any phase-difference offset near $\tau = 0$ because its strength is very weak in the vicinity of $\tau = 0$, see Fig. 2 (e). Our model also shows that all of the non-resonant current components reverse their direction with a change of the sign of detuning. This explains the inversion of the slope of the phase difference near $\tau = 0$ for a variation of the excitation photon energy across the lhX resonance, see Figs. 1 and 3. Although our experimental technique can be used to eliminate the contribution of the $IC_R$ to the measured TI, this does not hold for the $IC_{NR}$ whose temporal shape will be different. (This is because the real and imaginary parts of the modified electric field have different temporal shapes.) Since the TIs of $IC_{NR}$ and $ASC_R$ are the same, see Fig. 2 (d), it is most likely impossible to distinguish between these two current components for non-resonant excitation of discrete transitions. This visualizes the complexity of the overall current response that is already obtained for the simplest case, a single, discrete transition.

In conclusion, we have shown that additional photocurrent components arise from the non-resonant all-optical excitation of the fundamental light-hole exciton in a (110)-oriented GaAs quantum well. Ultrafast polarization-shaped optical pulses can be used to induce these currents, which only exist if the coherent polarization is noninstantaneous and frequency modulated. The currents vanish for CW excitation and thus reveal a new branch of photocurrents existing only in the ultrafast excitation regime. Among three possible non-resonant current components ($IC_{NR}$, $SSC_{NR}$, $ASC_{NR}$) we could experimentally only prove the existence of the $SSC_{NR}$; the other two components are either indistinguishable from other resonant currents or their amplitude is too small to be undoubtedly detected. We believe that our observations contribute to a better understanding of light-matter interaction since they highlight the importance of the noninstantaneous and frequency-modulated coherent response and show the complexity of current response tensors.

The authors thank Holger Marx for technical assistance and the Deutsche Forschungsgemeinschaft (DFG) for financial support.

**References:**

1. E. O. Göbel, K. Leo, T. C. Damen, J. Shah, S. Schmitt-Rink, W. Schäfer, J. F. Müller, and K. Köhler, "Quantum beats of excitons in quantum wells," Phys. Rev. Lett. **64**, 1801-1804 (1990).

2. J.-Y. Bigot, M.-A. Mycek, S. Weiss, R. G. Ulbrich, and D. S. Chemla, "Instantaneous frequency dynamics of coherent wave mixing in semiconductor quantum wells," Phys. Rev. Lett. **70**, 3307-3310 (1993).

3. D. S. Chemla, J.-Y. Bigot, M.-A. Mycek, S. Weiss, and W. Schäfer, "Ultrafast phase dynamics of coherent emission from excitons in GaAs quantum wells," Phys. Rev. B **50**, 8439-8453 (1994).

4. P. C. M. Planken, M. C. Nuss, I. Brener, K. W. Goossen, M. S. C. Luo, S. L. Chuang, and L. Pfeiffer, "Terahertz emission in single quantum wells after coherent optical excitation of light hole and heavy hole excitons," Phys. Rev. Lett. **69**, 3800-3803 (1992).

5. S. Priyadarshi, K. Pierz, U. Siegner, P. Dawson, and M. Bieler, "All-optical generation of coherent in-plane charge oscillations in GaAs quantum wells," Phys. Rev. B **83**, 121307(R) (2011).

6. X. Y. Yu, Q. Luo, W. L. Li, Q. Li, Z. R. Qiu, and J. Y. Zhou, "Ultrafast phase dynamics of coherent carriers in GaAs," Appl. Phys. Lett. **73**, 3321 (1998).

7. N. Sekine, Y. Shimada, and K. Hirakawa, "Dephasing mechanisms of Bloch oscillations in GaAs/Al$_{0.3}$Ga$_{0.7}$As superlattices investigated by time-resolved terahertz spectroscopy," Appl. Phys. Lett. **83**, 4794 (2003).

8. Y. Shimada, K. Hirakawa, M. Odnoblioudov, and K. A. Chao, "Terahertz Conductivity and Possible Bloch Gain in Semiconductor Superlattices," Phys. Rev. Lett. **90**, 046806 (2003).

9. I. D. Abella, N. A. Kurnit, and S. R. Hartmann, "Photon Echoes," Phys. Rev. Lett. **141**, 391 (1966).

10. M. Bieler, K. Pierz, U. Siegner, and P. Dawson, "Shift currents from symmetry reduction and Coulomb effects in GaAs/AlGaAs quantum wells," Phys. Rev. B **76**, 161304(R) (2007).

11. M. Bieler, K. Pierz, U. Siegner, and P. Dawson, "Quantum interference currents by excitation of heavy- and light-hole excitons in GaAs/AlGaAs quantum wells," Phys. Rev. B **73**, 241312(R) (2006).

12. S. Priyadarshi, A. M. Racu, K. Pierz, U. Siegner, M. Bieler, H. T. Duc, J. Förstner, and T. Meier, "Reversal of Coherently Controlled Ultrafast Photocurrents by Band Mixing in Undoped GaAs Quantum Wells," Phys. Rev. Lett. **104**, 217401 (2010).

13. J. E. Sipe and A. I. Shkrebtii, "Second-order optical response in semiconductors," Phys. Rev. B **61**, 5337 (2000).

14. S. Priyadarshi, K. Pierz, and M. Bieler, "All-optically induced ultrafast photocurrents: Beyond the instantaneous coherent response," Phys. Rev. Lett. **109**, 216601 (2012).

15. The 3$^{rd}$ rank tensors are denoted by three Cartesian indices, where the last two indices denote the polarization directions of the excitation fields and the first one denotes the direction of current flow.

16. L. E. Golub, "Spin-splitting-induced photogalvanic effect in quantum wells," Phys. Rev. B **67**, 235320 (2003).

17. V. I. Belinicher, E. L. Ivchenko, and B. I. Sturman, "Kinetic theory of the displacement photovoltaic effect in piezoelectrics," Sov. Phys. JETP **56**, 359 (1982).

18. M. Bieler, K. Pierz, and U. Siegner, "Simultaneous generation of shift and injection currents in (110)-grown GaAs/AlGaAs quantum wells," J. Appl. Phys. **100**, 83710 (2006).

19. G. Gallot and D. Grischkowsky, "Electro-optic detection of terahertz radiation," J. Opt. Soc. Am. B **16**, 1204 (1999).

20. Heavy-hole continuum transitions of the QW region and heavy-hole dominated continuum transitions of the bulk substrate are excited along with lhXs.

21. S. Schmitt-Rink and D. S. Chemla, "Collective Excitations and the Dynamical Stark Effect in a Coherently Driven Exciton System," Phys. Rev. Lett. **57**, 2752 (1986).

22. A. Nahata, A. S. Weling, and T. F. Heinz, "A wideband coherent terahertz spectroscopy system using optical rectification and electro-optic sampling," Appl. Phys. Lett. **69**, 2321-2323 (1996).

**Figure 1**

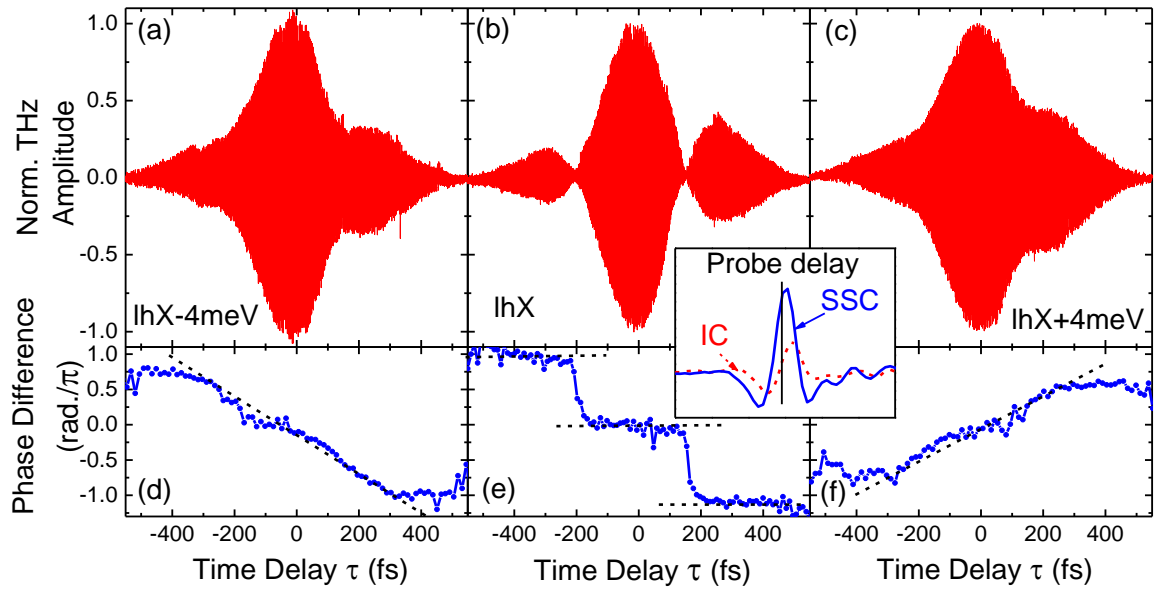

Fig. 1 (Color online) TIs for three different excitation photon energies: (a) ~4 meV below lhX, (b) at lhX, (c) ~4 meV above lhX. (d), (e), (f): Corresponding phase differences between the EFC and the TI. Inset: Temporal THz traces of SSC and IC for excitation of the lhX. The vertical solid line shows the temporal position of the probe beam used to obtain the TI and this position has been varied by τ/2 to account for a change in τ. The dotted lines are guides for the eye.

**Figure 2**

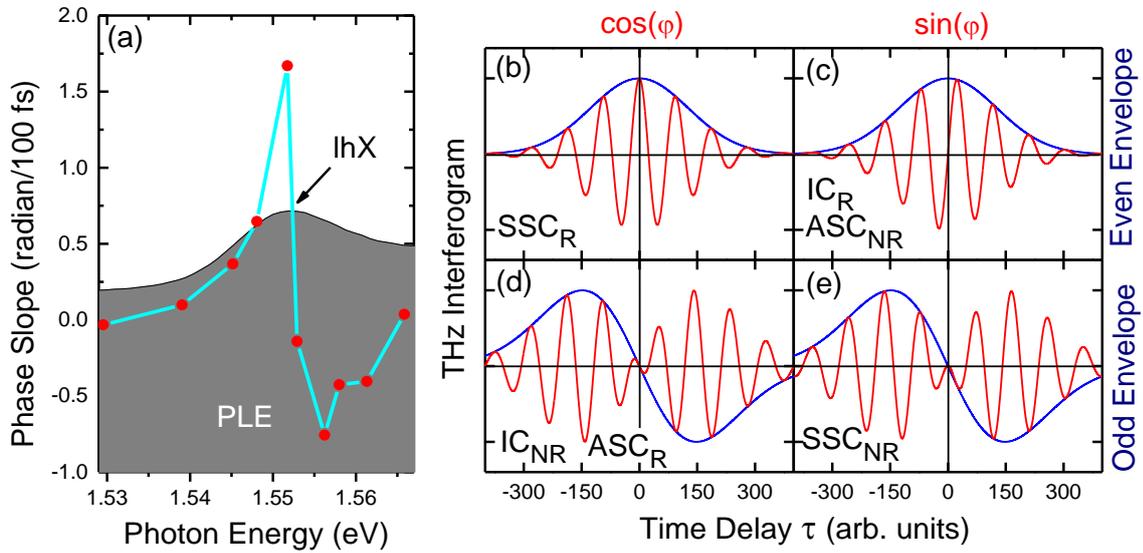

Fig. 2 (Color online) (a) Slope of the phase difference versus excitation photon energy. Gray background shows photoluminescence excitation spectrum (PLE) of the QW sample. Illustration of TIs for (b) $SSC_R$ with an even envelope and $\cos(\varphi)$ dependence, (c) $IC_R$ and $ASC_{NR}$ with an even envelope and $\sin(\varphi)$ dependence, (d) $IC_{NR}$ and $ASC_R$ with an odd envelope and $\cos(\varphi)$ dependence, and (e) $SSC_{NR}$ with an odd envelope and $\sin(\varphi)$ dependence. For illustration purposes the oscillation period determined by the $\cos(\varphi)$ and $\sin(\varphi)$ functions has been considerably increased as compared to the width of the envelopes.

**Figure 3**

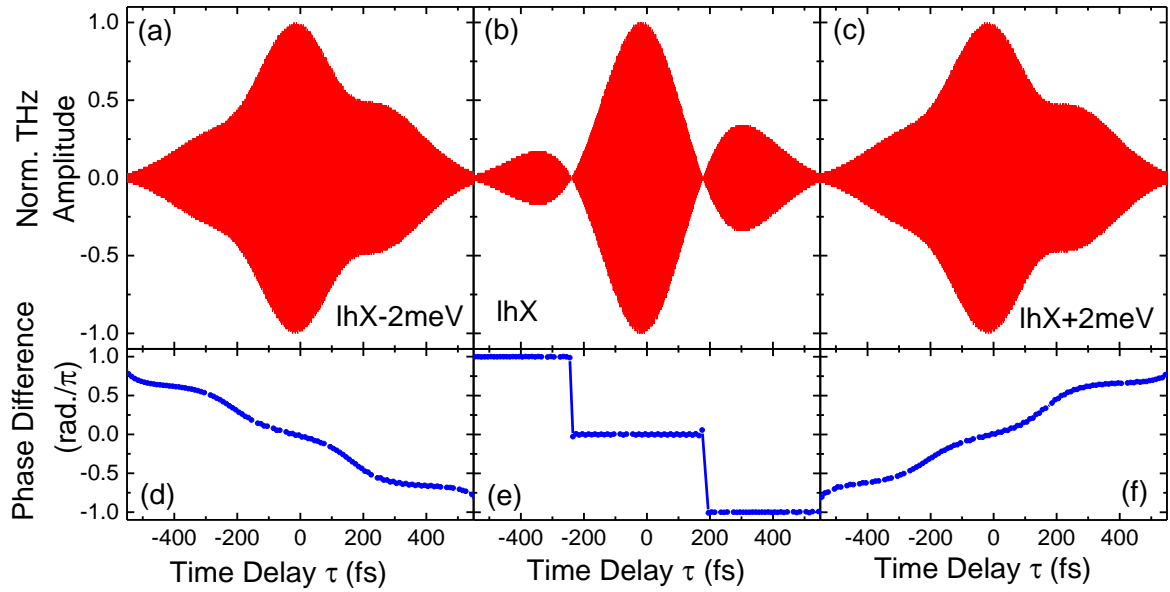

Fig. 3 (Color online) Simulation of TIs for three different excitation photon energies: (a) ~2 meV below lhX, (b) at lhX, (c) ~2 meV above lhX. The corresponding phase differences are shown in (d), (e), and (f). We obtain the best agreement with the experiments for smaller detuning in the simulations (2 meV instead of 4 meV). Most likely this is because our simple model does not account for inhomogeneous broadening.